# Computational study of antiferromagnetic and mixed-valent diamagnetic phase of $AgF_2$: crystal, electronic and phonon structure and p-T phase diagram

**Kamil Tokár,[a;b] Mariana Derzsi, [a;c] and Wojciech Grochala[c]**

Crystal and electronic structure, lattice dynamics and thermodynamic stability of little known mixed-valent diamagnetic $Ag^{1+}Ag^{3+}F_4$ β form of $AgF_2$ is thoroughly examined for the first time and compared with well known antiferromagnetic $Ag^{2+}F_2$ α form within the framework of Density Func-tional Theory based methods, phonon direct method and quasiharmonic approximation. Computed equations of state, bulk moduli, electronic densities of states, electronic and phonon band structures including analysis of optically active modes and p–T phase diagram of the α–β system are presented. This study demonstrates that α is thermodynamically preferred over β at all temperatures and pressures of its existance but simultaneously b is dynamically stable in much broader pressure range. The β phase is discussed in broader context of isostructural ternary metal fluorides and isolectronic oxides including $NaCuO_2$ – the reference compound for existence of $Cu^{3+}$ species in high-temperature oxocuprate superconductors.

## 1 Introduction

Two polymorphic forms have been reported for $AgF_2$ at ambi-ent conditions – layered antiferromagnetic (α) and diamagnetic (β). The a phase exhibits numerous structural and electronic similarities with oxocuprate precursors of high-temperature su-perconductors, including the 2D antiferromagnetic (AFM) $AgF_2$ sheets with large AFM super-exchange constant that have the same topology as [$CuO_2$] layers that mediate superconductivity in the high-temperature superconducting oxocuprates. [1] Physico-chemical properties of the a phase including crystal and magnetic structure [2–4], electronic structure, [1,5] and high-pressure polymorphism [6–9] have been thoroughly scrutinized and a phase is a sub-ject of many of our ongoing studies. On the other hand, the β phase is much less understood. It has been obtained as a red-brown diamagnetic product of reaction of $AgBF_4$ with $KAgF_4$ in anhydrous HF and magnetic measurements suggested mixed-valent $Ag^{1+}Ag^{3+}F_4$ system. [10] Although its crystal structure could not be determined due to the lack of crystallinity it is assumed to take $KBrF_4$ structure type, which is typical for $Ag^{1+}Au^{3+}F_4$ and other ternary metal fluorides $M'MF_4$ with coinage (M) and alkali metals (M'). [10–12] Notably, the β phase undergoes a rapid exothermic conversion to the a form accompanied with charge transfer $Ag^{1+}Ag^{3+}F_4$ ! $2Ag^{2+}F_2$ when the temperature is raised from -80 C to 0 C. This behaviour can be contrasted with charge-transfer instability in oxocuprates. Some authors suggest that charge-transfer instability of the parent insulating cuprates may drive their unconventional superconductivity. [13] A direct relation between unconventional superconductivity and the dispro-portionation reaction is well documented for charge-ordered in-sulator $BaBi^{3+/5+}O_3$, which can be converted to a superconductor by a nonisovalent substitution $Ba_{1\ x}K_xBiO_3$. However, the charge ordered $Cu^{1+}/Cu^{3+}$ form has never been observed in undoped oxocuprates. Instead, $NaCuO_2$ has been studied as reference compound for the existence of $Cu^{3+}$ species in high-temperature oxocuprate superconductors. [14–16] On the other hand, its silver counterpart in form of $Ag^{1+}Ag^{3+}O_2$ is known but in this case the comproportionated $Ag^{2+}O$ has never been observed. [17–22] In-stead it exists in two disproportionated polymorphic forms, mon-oclinic [17] and tetragonal, [18] and remains mixed-valent semicon-ductor up to extremely high pressures (75 GPa). [23] Thus, the ap-pearance of both the comproportionated $Ag^{2+}$ and disproportion-ated $Ag^{1+/3+}$ states in $AgF_2$ renders this system unique and re-quires detailed examination.

[a] *Advanced Technologies Research Institute, Faculty of Materials Science and Technology in Trnava, Slovak University of Technology in Bratislava, J. Bottu 25, 917 24 Trnava, Slovakia. E-mail: mariana.derzsi@stuba.sk*

[b] *Institute of Physics, Slovak Academy of Sciences, Dúbravská cesta 9, 845 11 Bratislava, Slovakia.*

[c] *Center of New Technologies, University of Warsaw, Zwirki i Wigury 93, 02089 War-saw, Poland.*

Crystal structure and stability of the β relative to the α $AgF_2$ were first addressed in a theoretical Density Functional Theory (DFT) study dedicated to pressure-induced transformations of $AgF_2$ [6]. The study revealed metastability of the β over a by +0.17 eV per one $AgF_2$ unit at ambient conditions and growing further with increasing pressure. However, only results of total energy calculations were presented while the impact of lattice dynamics and temperature have remained open questions; the former was later demonstrated to have crucial impact on high-pressure phase transitions in the α phase. [8,9] The study also did not consider the strong electron correlations and magnetism, which are important for proper description of $AgF_2$ system. [1,4,5,8,9] It is also not known if the $KBrF_4$ type candidate for the β phase is dynamically sta-ble and if so, what is the character of its electronic structure and how do the crystal, electronic and phonon structure evolve with pressure. Considering the a phase, although it is much better understood, its lattice dynamics has not been analyzed in detail before and impact of temperature on its stability is also unknown. In this work, we address all the above issues in a computational study from the perspective of DFT-based methods, phonon direct method and quasiharmonic approximation while accounting for strong electron correlations and magnetism.

Here we present results of a detailed comparative computa-tional DFT+U study of crystal and electronic structure and lattice dynamics of the α and the β $AgF_2$ phase (with larger focus on the less known b , considering the $KBrF_4$ structure type) and a theo-retical p–T phase diagram of the α β system. The crystal, elec-tronic and phonon structure of both phases are compared also as functions of pressure and pressure-dependences of their Raman and IR active frequencies are provided. Furthermore, we discuss the electronic structure of both phases from the perspective of DFT+U and hybrid DFT method since the latter is known to pro-vide more realistic values for the insulating band gaps. [24] Finally, all our results obtained for the β phase are confronted with avail-able literature data for isostructural fluorides $M'MF_4$ with coinage (M) and alkali metals (M') and with isoelectronic $NaCu^{3+}O_2$ and $Ag^{1+}Ag^{3+}O_2$.

## 2 Calculation details

Periodic DFT calculations for the a and b phase of $AgF_2$ were performed in the plane-wave VASP program with PAW-GGA method, [25,26] PBEsol functional, [27] plane-wave cut-off energy set to 520 eV and k-mesh spacing of 0.28 Å [1] for cell relaxation and 0.14 Å [1] for electronic structure calculations. The onsite electron correlations on Ag d orbitals were accounted for using the ro-tationally invariant DFT+U method introduced by Liechtenstein *et al.*, where the values of both Hubbard U and the Hund J parameter are set explicitly [28]. We have used the values $U_{Ag}$ = 5 eV and $J_{Ag}$ = 1 eV in accordance with the previous studies. [4,8,9] In case of the a phase, spin-polarized calculations were per-formed considering the known AFM ground-state. [2–4] Electronic density of states and band structure were further recalculated for the DFT+U optimized structures with hybrid HSE06 func-tional with 25% of exact exchange. [29] Phonon dispersion curves and phonon density of states (PDOS) were calculated with direct phonon method in the pressure interval ranging from -6 GPa to +8 GPa (at higher pressures the a phase transforms to a ferroelec-tric $Pca2_1$ form [8]). These calculations were performed for 2 x 2 x 2 (a) and 2 x 2 x 1 (b ) supercells (96 atoms each), with the recip-rocal space sampled over the 4 x 4 x 4 Monkhorst–Pack k-mesh. [30] The pressure dependent PDOSes were subsequently used to ob-tain thermodynamic functions within quasi-harmonic approxima-tion (QHA). The phonon and thermodynamics calculations were performed in program PHONOPY. [31] The force constants were de-rived from Hellmann-Feynman forces that were computed with finite atomic displacement equal to 0.02 Å at DFT+U level. Dy-namical matrices were constructed considering full symmetries of the supercell. The equation of state obtained in QHA approxima-tion enables for estimation of thermoelastic properties. The a b phase diagram was constructed from the Gibbs energy isobars in the p–T diagram. Thermodynamic potentials of both crystalline phases were derived from the free energy $F_{phon}(T, V)$ of calcu-lated phonon system and ground state energies of the phases at given external pressure p and temperature T within QHA. Phase relations between a and b phases and their relative thermody-namic stability were constrained from Gibbs energy isobars by determining the boundary locus of points in p-T phase diagram, where crystal structures of the two phases satisfy Gibbs rule of thermal equilibrium $G^a$ (T,p) = $G^b$ (T,p).

## 3 Results

### 3.1 Crystal structure

Crystal structures of a and b $AgF_2$ are shown in Figure 1.

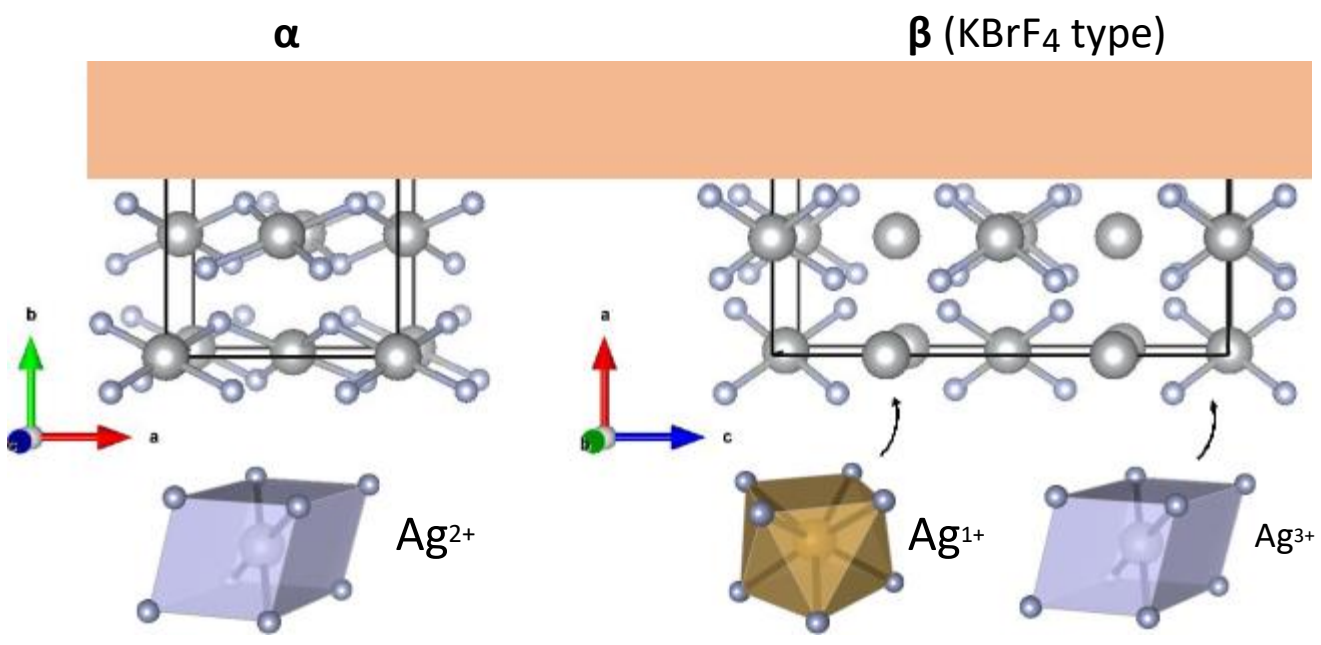

Fig. 1 Comparison of crystal structure of α and β $AgF_2$ and their ex-tended Ag coordination polyhedra: (left) experimental orthorhombic a with Pnma symmetry and (right) hypothetical tetragonal β assuming $KAgF_4$ I4/mcm structure type. Note, β can be viewed as a structure, where every other $Ag^{2+}$ F bond within each $AgF_2$ layer (highlighted with orange stripe) has been broken. The bonds within the coordination polyhedra indicate the first coordination sphere.

The a phase crystallizes in orthorhombic Pbca symmetry with neutral $AgF_2$ layers stacked in ABAB fashion along b crystallo-graphic axes. Coordination of the $Ag^{2+}$ ions within the layers is square planar [$AgF_4$]. The first nearest-neighbour axial F con-tacts complete elongated octahedral coordination and the sec-ond nearest-neighbour ones complete deformed rhombic prism (2+2+2+2 coordination). The structure was thoroughly char-acterized before. [2] Our DFT+U calculations reproduce very well all experimental structural parameters including Ag F distances (the short square-planar distances within 0.001 Å, the longer axial contacts within 0.02 Å) and unit cell volumes (within less than 1%) (Table 1).

Table 1 Selected structural data for α- and β -$AgF_2$ and for $A^{1+}Ag^{3+}F_4$ and $A^{1+}Au^{3+}F_4$ compounds (where A stands for alkali metal) with $KBrF_4$ type structure. DFT+U values were computed in this work and experimental values are taken from literature.

| Struc. | Ref. | V/$AgF_2$ | a | b | c | $M^{2+}$–$F_{eq}$ | $M^{2+}$–$F_{ax}$ | $M^{3+}$–$F_{eq}$ | $M^{3+}$–$F_{ax}$ | $Ag^{1+}$–F | F–$Ag^{3+}$–F | F–$Ag^{1+}$–F |
|---|---|---|---|---|---|---|---|---|---|---|---|---|
| a-$AgF_2$ | exp [2] | 40.8 | 5.529 | 5.813 | 5.073 | 2.067, 2.071 | 2.588 | | | | | |
| a-$AgF_2$ | DFT+U | 40.5 | 5.498 | 5.824 | 5.055 | 2.069, 2.071 | 2.568 | | | | | |
| b -$AgF_2$ | DFT+U | 40.8 | 5.524 | | 10.704 | | | 1.933 | 2.848 | 2.432 | 92.2 | 113.4 |
| $NaAgF_4$ | exp [32] | 40.5 | 5.540 | | 10.560 | | | 1.896 | 2.906 | 2.430 | 89.9 | 115.4 |
| $AgAuF_4$ | exp [33] | 45.4 | 5.791 | | 10.817 | | | 1.885 | 3.016 | 2.577 | 93.3 | 113.6 |
| $KAgF_4$ | exp [32] | 48.5 | 5.900 | | 11.150 | | | 1.896 | 3.129 | 2.646 | 90.2 | 113.6 |
| $KAuF_4$ | exp [34] | 51.0 | 5.990 | | 11.380 | | | 1.998 | 3.154 | 2.652 | 90.1 | 114.5 |
| $RbAuF_4$ | exp [32] | 56.6 | 6.180 | | 11.850 | | | 1.998 | 3.273 | 2.788 | 90.2 | 112.3 |

In β phase with tetragonal $KBrF_4$ structure two non-equivalent Ag ions are present, square planar and square antiprismatic one organized into separate layers alternating along the tetragonal c direction (Figure 1). The [$AgF_4$] square plaquettes are isolated from each other, oriented perpendicular to the ab layers and to each other in all three crystallographic directions. The calcu-lated Ag F distances of the square-planar silver (1.933 Å) are noticeably shorter relative to the corresponding distances in a (2.069 and 2.071 Å). This is in line with slightly smaller $Ag^{3+}$ ion in respect to $Ag^{2+}$ and indeed the values follow closely those expected based on tabulated ionic radii by Shannon, [35] $Ag^{3+}$ F (1.98 Å) and $Ag^{2+}$ F (2.1 Å) ones. The square-planar silver in β has additionally four secondary axial contacts equal to 2.848 Å that complete its coordination to distorted rhombic prism. This contrasts with much shorter axial contacts equal to 2.568 Å that complete the octahedral coordination of $Ag^{2+}$ in a. The longer secondary contacts in b reflect the tendency of $Ag^{3+}$ to avoid mutual secondary interactions $Ag^{3+}$F. . . $Ag^{3+}$. From this point of view, $KBrF_4$ type structure is an ideal host for $Ag^{3+}$ since it effectively minimizes the mutual secondary $Ag^{3+}$F. . . $Ag^{3+}$ inter-actions by allowing for perfect antiferrodistortive ordering of the [$Ag^{3+}F_4$] plaquettes in all three crystallographic directions. The square antiprismatic silver has eight equivalent Ag F distances equal to 2.432 Å. This value is slightly shorter than the one ob-served for $Ag^{1+}$ in isostructural $Ag^{1+}Au^{3+}F_4$ (2.577 Å), [33] which in turn is very close to the value based on the tabulated ionic radius for 8-coordinated $Ag^{1+}$ (2.59 Å). On the other hand, it is almost identical to the $Na^{+}$ F distance (2.43 Å) in isostruc-tural $Na^{1+}Ag^{3+}F_4$. [32] In fact, all structural parameters calculated for the β phase follow very closely those of $NaAgF_4$ including almost identical volumes of 326.7 ($Ag^{1+}$) and 324.1 ($Na^{1+}$) and primary/secondary $Ag^{3+}$ F contacts of 1.933 Å /2.848 Å (in $Ag^{1+}Ag^{3+}F_4$) and 1.896 Å /2.906 Å (in $Na^{1+}Ag^{3+}F_4$). These re-sults are in line with the well-known similarities between $Ag^{1+}$ and $Na^{1+}$. Comparison with other isostructural compounds con-taining $Ag^{3+}$ and $Au^{3+}$ species may be observed in Table 1.

Despite the above described differences between α and β form, their crystal structures are intimately related. Both share the same fluorite metal sublattice and differ only in different displacements of the fluorine atoms. In case of the β phase, the displacements of F atoms lead to doubling of the fluorite unicell along the c axis. Also note, that one can rationalize the β phase as the a phase with bond-broaked $AgF_2$ layers (Figure 1). The breaking of the Ag-F bonds takes place due to charge disproportionation $Ag^{2+}Ag^{2+}$ ! $Ag^{1+}Ag^{3+}$ and leads to doubling of the a lattice vector of α (Figure 1). It should be noted that in the fluorite struc-ture, the metal cations are in cubic coordination with the ligands. In the a phase, the extended coordination of silver cations is in first approximation angularly deformed cubic (rhomboid), while the orthorhombic symmetry drives the coordination to an elon-gated octahedron. In the β phase, half of the coordination poly-hedra belonging to $Ag^{3+}$ are deformed rhomboids and another half ($Ag^{1+}$) are square antiprisms or in another words twisted cubes. In fact, both structures are related to the fluorite structure by group-subgroup relation involving in each case single order parameter X5+, resp. L3+/L3-: Fm-3m (fluorite)! $^{X5+}$ Pbca (a) and Fm-3m (fluorite)! $^{L3+(L3-)}$ I4/mcm (b ) and thus could be understood as emerging from the fluorite structure via a distinct ordering parameter. The group-subgroup analysis was performed using ISOTROPY Software Suite. [36]

Quite remarkably, the computed volumes per one $AgF_2$ unit of α and β are extremely similar. The calculated unit cell volumes at zero external pressure are $V_a$ = 161.86 Å$^3$ and $V_b$ = 326.66 Å$^3$ (they contain 4 and 8 $AgF_2$ units per unit cell, respectively). These translate to 40.47 Å$^3$ and 40.83 Å$^3$ per one $AgF_2$ unit, re-spectively; β having less than 1% larger volume. Interestingly, this relationship holds also under pressure at least up to 10 GPa (Figure 2). Around this pressure, the α form transforms to a non-centrosymmetric $Pca2_1$ structure. [8] Similar compressibility of the two forms translates to bulk modulus of b ($B_0$ = 54 GPa) being only slightly larger than bulk modulus of a ($B_0$ = 50 GPa). Interestingly, compressibility of both forms is comparable despite their entirely different electronic character as discussed in Section 3.2. It is simultaneously much higher when compared to other electronically and structurally related compounds. For example, $CuF_2$, which is the lighter counterpart of a-$AgF_2$ is considerably less compressible with DFT+U bulk modulus equal to 75 GPa. [37] Additionally, when computed in α-$AgF_2$ structure type, its bulk modulus amounts to 71 GPa. [37] Since both, $CuF_2$ and α-$AgF_2$ are isolectronic and their crystal structures differ only in stacking of otherwise similar puckered metal fluoride layers, the larger compressibility of a-$AgF_2$ is related to presence of larger (and softer) $Ag^{2+}$ cation relative to $Cu^{2+}$. In case of β-$AgF_2$, neither its lighter nor heavier counterpart exists. The only isoelectronic compound with known compressibility is $Ag^{1+}Ag^{3+}O_2$. Its bulk modulus es-timated from high-pressure X-ray diffraction and DFT+U data is even higher close to 83 GPa. [23] However, its crystal as well as electronic structure differ from β -$AgF_2$ as discussed in Section 3.5.

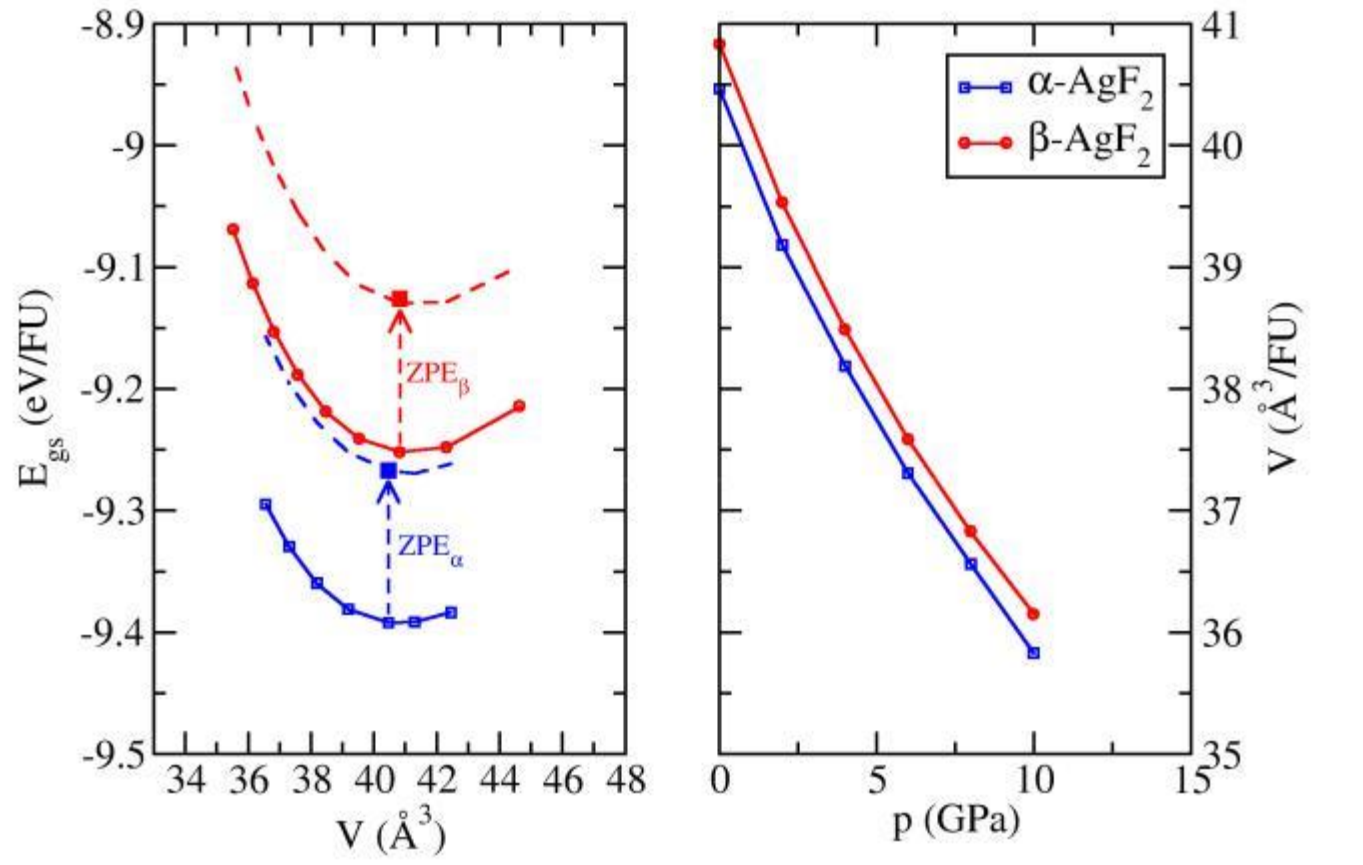

Fig. 2 Equation of state for α- and β -$AgF_2$ for T=0 K (DFT+U re-sults). The left panel depicts the static energy dependence on volume per one $AgF_2$ unit (continuous curves) together with ZPE correction (dashed curves) plotted with a 3rd-order polynomial fit. The energy minima correspond to zero external pressure. The right panel depicts p-V dependence.

The β phase is the most compressible within the tetragonal plane. At 8 GPa the a and c lattice vectors reduce to 95.6% and 98.4% of their zero-pressure values, respectively. The higher com-pressibility within the tetragonal plane is governed by the four axial $Ag^{3+}$F. . . F contacts. They are the longest and the most com-pressible contacts in the structure running along the ab plain diag-onal. At 8 GPa, they experience reduction by 0.13 Å from 2.848 Å to 2.719 Å. Second in line are the eight square antiprismatic $Ag^{1+}$ F contacts that reduce by 0.082 Å from 2.432 Å to 2.331 Å. The least compressible or the most rigid ones are the square planar $Ag^{3+}$ F bonds. They reduce only by 0.004 Å from 1.933 Å to 1.927 Å. Even at pressure of 20 GPa, the length of these bonds is reduced by no more than 0.018 Å in contrast with 0.224 Å and 0.373 Å reduction for the $Ag^{1+}$ F and $Ag^{3+}$F. . . F respec-tively, which further manifests the rigidity of the former bonds. The rigidity of the square-planar $Ag^{3+}$ F bonds is comparable to that of the square-planar $Ag^{2+}$ F bonds in the α phase. The lat-ter was previously examined in a combined high-pressure XRD and DFT study. [8] The lattice parameters of α and β computed in function of pressure are compared in Table S1 in ESI.

It is interesting to observe evolution of the tetragonal c/a ra-tio of the β phase under pressure and confront it with the range of c/a ratios observed in related compounds with $KBrF_4$ struc-ture congaing coinage metals. The c/a ratio of the β phase at zero pressure is 1.94 and it increases with pressure. For $Ag^{3+}$ and $Au^{3+}$ compounds this ratio is found within the range 1.87 – 1.92, [12,32–34] while the highest value of 2.06 is observed for $CsCu^{3+}F_4$. [38] This seems to be the limiting case, since $LiAuF_4$ and $CsAuF_4$ with higher c/a rations of 2.07 and 2.10 crystallize in mon-oclinic and orthorhombic variant of $KBrF_4$ structure, respectively. According to our DFT+U results, b phase would reach this value at pressures above 20 GPa. This result can be confronted with our phonon calculations (discussed below), which indicate dynami-cal stability at least up to 20 GPa, the highest calculated pressure. Interestingly, $KBrF_4$ type compounds containing $Ag^{2+}$ in combina-tion with divalent cations such as $Ca^{2+}$, $Sr^{2+}$, $Ba^{2+}$, $Zn^{2+}$, $Cd^{2+}$ and $Hg^{2+}$ instead of $K^+$ are also known and their c/a ratios are observed within the range 1.9 – 1.99. [39] Note, this ratio is slightly higher than for the $KBrF_4$ type compounds with $Ag^{3+}$ (1.89 for $NaAgF_4$ and 1.91 for $KAgF_4$). In β , the highest of these two val-ues is reached at 8 GPa. This observation alone suggests that b phase has the potential to accommodate also $Ag^{2+}$ in a wide pressure interval. Since the $KBrF_4$ type compounds with $Ag^{2+}$ are paramagnetic, [39] we have tried to reach magnetic solutions for the β phase at all calculated pressures. However, our attempts to stabilize antiferromagnetic solutions for β were unsuccessful and ferromagnetic solution was found higher in energy in respect to diamagnetic at all pressures.

### 3.2 Electronic structure

Electronic structures of a and β -$AgF_2$ are presented in Figure 3 (electronic DOS) and Figure 4 (electronic band structure).

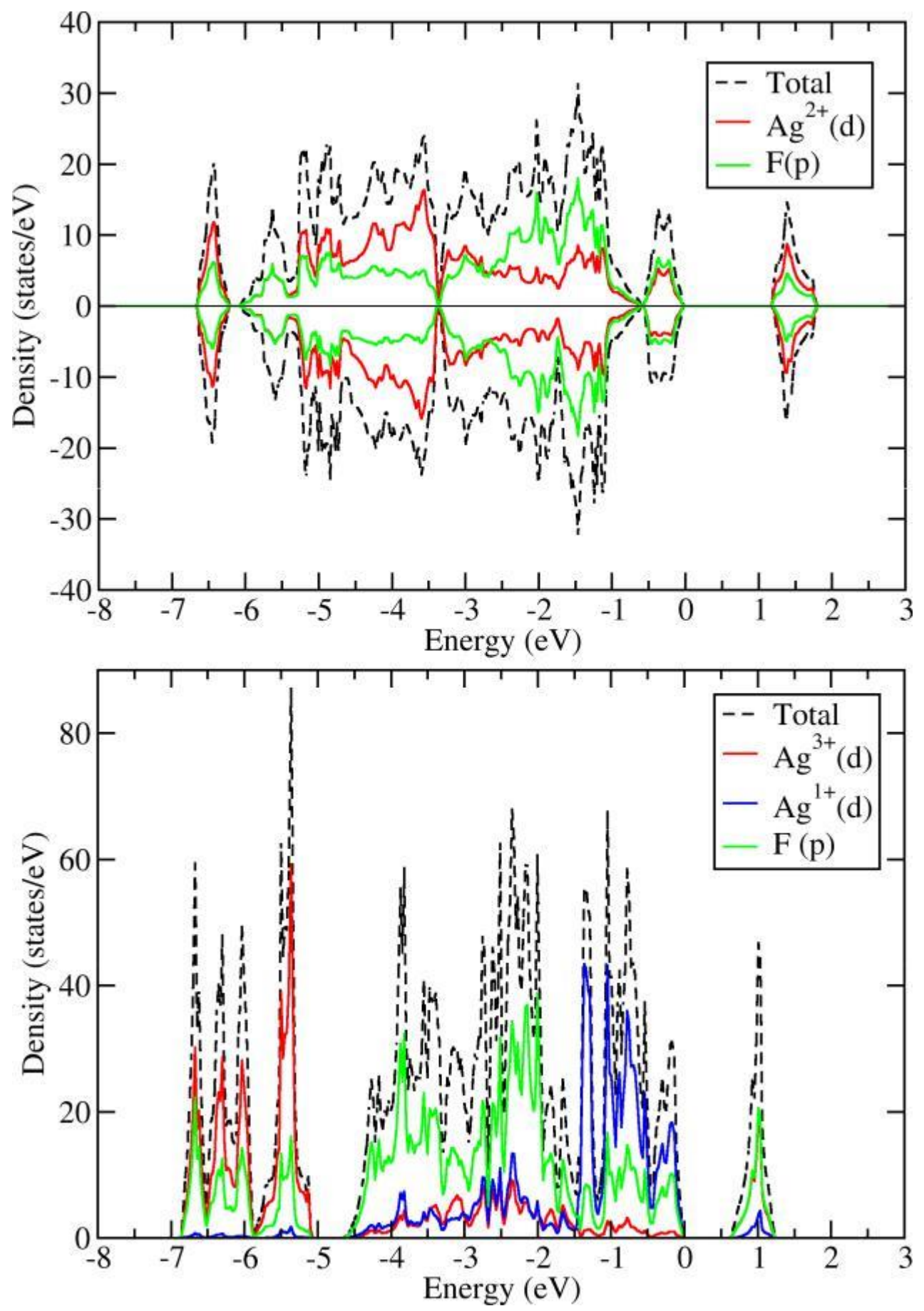

Fig. 3 Electronic DOS of antiferromagnetic α (top) and diamagnetic β $AgF_2$ (bottom) calculated with DFT+U.

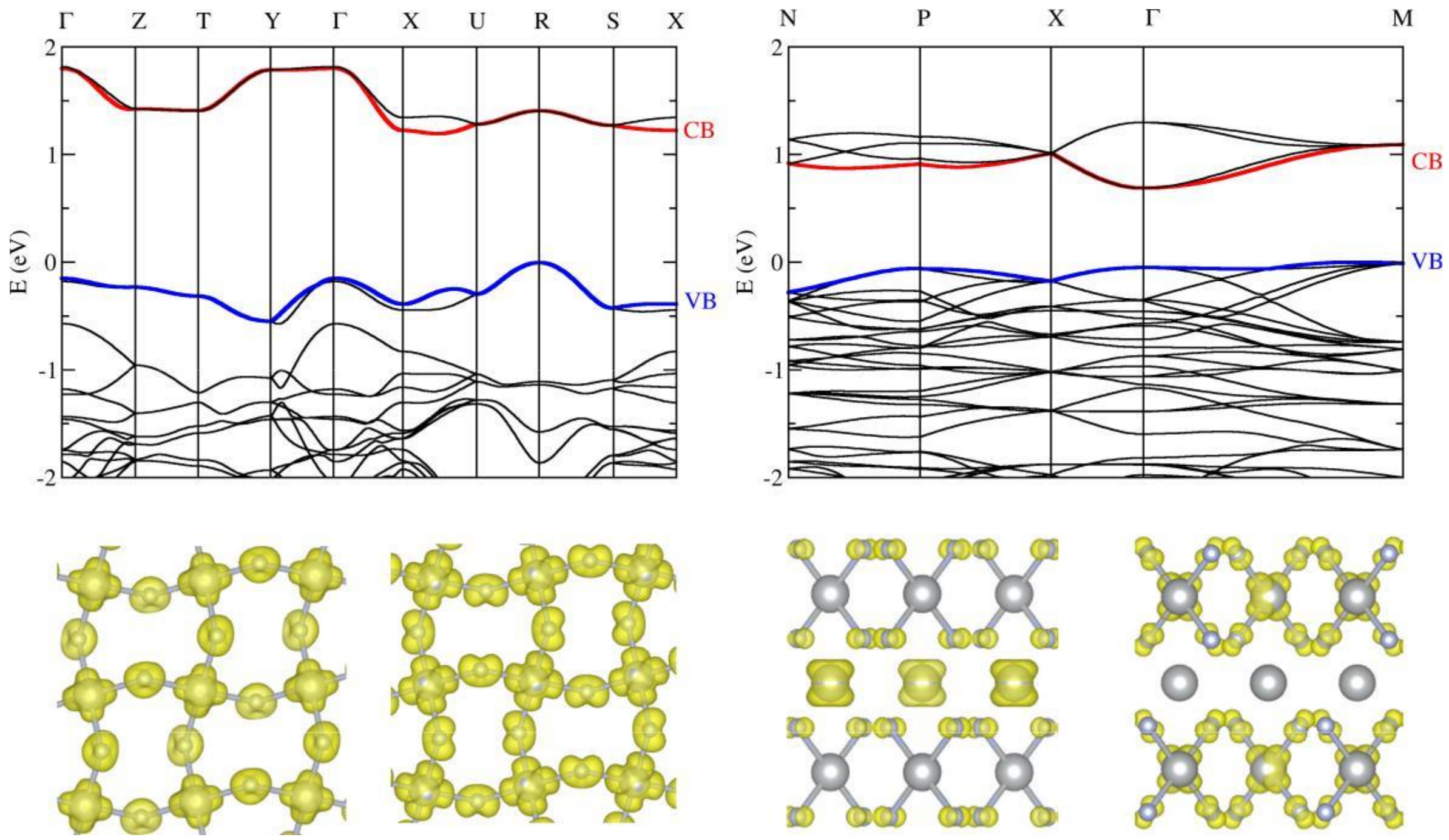

Fig. 4 Electronic band structure (BS, top) and real-space projection of the valence band (VB, blue curve in the BS) and the conduction band (CB, red curve in the BS) (bottom) of α- (left) and β -$AgF_2$ (right) calculated with DFT+U.

Electronic density state (eDOS) of α-$AgF_2$ projected on individual atoms reveal strong mixing of Ag and F states across the entire 7 eV wide valence region (Figure 3, top). The large overlap of Ag and F states is a consequence of the extended 2D network of short Ag-F bonds. The highly localized covalent character of the Ag-F bonds is witnessed by the narrow peaks positioned at the very bottom and very top of the valence DOS. The bonds form between the Ag d($x^2$ -$y^2$) and F p(x)/p(y) orbitals. The bottom peak cen-tred at 6:5 eV corresponds to the respective s bonding states of prevalent Ag d($x^2$ -$y^2$) character (the lower Hubbard band) and the top one centred at 0:25 eV to s antibonding of predominant F p character. An insulating bandgap of 1:17 eV opens between the s antibonding states and upper Hubbard band of predomi-nant Ag d($x^2$ -$y^2$) character as expected for formally $Ag^{2+}$ cation with $d^9$ electronic configuration and a hole in the $e_g$ states. The real-space (orbital) projections of these bands reveal contribution from the p(z) states to the valence band but not to the conduction band (Figure 4, bottom left). Electronic band structure calculations reveal formation of indirect insulating bandgap along R – X direction in the Brillouin zone.(Figure 4, top left) These results show that α-$AgF_2$ is a charge-transfer indirect bandgap insulator in agreement with our previous studies. [1,5] Very similar electronic structure is observed in undoped cuprates, which makes $AgF_2$ an excellent cuprate analogue. [1]

A distinct picture of electronic structure is obtained for β -$AgF_2$. Here, the eDOS projected on individual atoms reveal distinct contribution from the square-planar and the square-antiprismatic sil-ver atom (Figure 3, bottom). The d states of the former are found predominantly in the bottom of the valence DOS and the d states of the latter silver are dominantly located in the upper parts of the valence DOS. In case of the square-antiprismantic silver, all d states are occupied as expected for closed shell $d^{10}$ $Ag^{1+}$ cation. On the other hand, the s antibonding d($x^2$ -$y^2$) states are de-populated in case of the square-planar silver in line with low spin $d^8$ configuration characteristic for $Ag^{3+}$ species involved in a co-valent bond. The s bonding $Ag^{3+}$ states are found at the very bottom of the valence DOS centered at 6.7 eV. However, the en-ergy separation of the s bonding and s antibonding states is larger in β ($\Delta_{\sigma\text{-}\sigma^*}$ = 7.6 eV) by 1.3 eV, which confirms pres-ence of even stronger covalent bonds in comparison to a ($\Delta_{\sigma\text{-}\sigma^*}$ = 6.25 eV). Furthermore, all $Ag^{3+}$ d states are much more lo-calized compared to $Ag^{2+}$ d states in a. They form highly lo-calized bands within narrow 3 eV–wide range between 7 and 5 eV separated from remaining valence DOS by a bandgap of 0.5 eV in contrast to $Ag^{2+}$ d states in a, which overlap with F p states across the entire valence DOS. The high localization of the $Ag^{3+}$ d states in b shows that they are not involved in forma-tion of extended bonding network in contrary to corner-sharing [$Ag^{2+}F_4$] plaquettes in a. Instead, they are confined to the co-valent bonds within the isolated [$Ag^{3+}F_4$] plaquettes held in the crystal by electrostatic interactions with $Ag^{1+}$ cations. Presence of typical $Ag^{1+}$–F bonds in β -$AgF_2$ is witnessed by highly local-ized character of the valence $Ag^{1+}$ states positioned on top of the valence F bands. A bandgap of 0.62 eV opens between the nonbonding $Ag^{1+}$ states and σ antibonding $Ag^{3+}$ states ($Ag^{1+}$ → $Ag^{3+}$ inter-valence charge-transfer) between M and G point of the Brillouin zone (Figure 4, top right). The real-space projec-tions of the highest occupied and the lowest unoccupied states show that the former is constructed from $Ag^{1+}$ d(xz), d(yz) and F p(z) atomic orbitals whereas the latter involves $Ag^{3+}$ d($x^2$ $y^2$), F p(x) and p(y) (Figure 4, bottom right). Thus, β -$AgF_2$ is a distinct charge-transfer insulator with the insulating bandgap by factor of 2 smaller compared to the insulating bandgap of the α phase.

Under compression, electronic structure of both $AgF_2$ phases does not change significantly within the pressure range of their dynamical stability. Both experience typical broadening of all electronic bands with increasing pressure. However, in case of the β phase, slow progress towards increased hybridization of all Ag d and F p states with the pressure (reminiscent of the a phase) is evident. Evolution of electronic DOS with pressure can be observed for both $AgF_2$ phases in Figures S2 and S3 in ESI. The band gaps of both phases decrease under compression while the band gap of a remains broader and both phases remain insulating at all calculated pressures. We have recalculated the band gaps also with hybrid DFT, which is known to provide a better estimate for insulating bandgaps. These calculations provide the zero-pressure value of 2.34 eV for α and 1.17 eV for β and so by factor of 2 wider bandgaps in respect to DFT+U. However, also on hybrid DFT level the bandgap of β is by factor of 2 smaller relative to a and retains its insulating character at all calculated pressures (Table 2). Similar DFT+U and hybrid DFT values of the band gap were obtained for the α phase in previous studies. [1,4,5] For completeness, evolution of electronic DOS calculated for α and β phase with hybrid DFT can be observed in Figures S4 and S5 in ESI. Similar picture is obtained as with DFT+U method in-cluding large overlap between $Ag^{2+}$ d and F p states for a and large localization of the $Ag^{3+}$ and $Ag^{1+}$ d at the bottom and top of the valence region for b . Additionally, Hybrid DFT points out to a larger mixing of the F p states with the $Ag^{3+}$ d and a smaller overlap with the $Ag^{1+}$ d states relative to DFT+U picture.

Table 2 Evolution of insulating bandgap with pressure calculated for the α- and β -$AgF_2$ on DFT+U and hybrid DFT (HSE06) level.

| | $E_{gap}{}^{a}$ | | $E_{gap}{}^{b}$ | |
|---|---|---|---|---|
| p(GPa) | DFT+U | HSE06 | DFT+U | HSE06 |
| 0 | 1.169 | 2.341 | 0.613 | 1.171 |
| 2 | 1.141 | 2.289 | 0.501 | 1.050 |
| 4 | 1.111 | 2.261 | 0.426 | 0.987 |
| 6 | 1.098 | 2.251 | 0.367 | 0.903 |
| 8 | 1.104 | 2.220 | 0.308 | 0.841 |

### 3.3 Lattice dynamics

The phonon dispersion curves and phonon density of states (PDOS) of the α and β phase integrated over entire Brillouin zone are shown jointly in Figure 5. Comparing first the phonon dispersion curves, one can immediately notice large difference in lattice dynamics between the two phases. In the β phase they cover broader frequency range 0–590 cm $^{1}$ in contrast to 0–500 cm $^{1}$ in α. Furthermore, the phonon bands of α are much more dispersed and coupled than those in the b phase.

In α, the phonons are split to two energy regions separated at around 200 cm $^{1}$ by a 25 cm $^{1}$ wide gap (Figure 5, top). The PDOS reveals that the higher energy region is dominated by con-tributions from F atoms, while Ag atoms contribute considerably to the bands in the range 0–200 cm $^{1}$ and dominate the phonon spectrum below 100 cm $^{1}$. The most dispersed bands are found in the higher energy region that covers the bond stretching and bending vibrations. The high dispersion of these bands results in

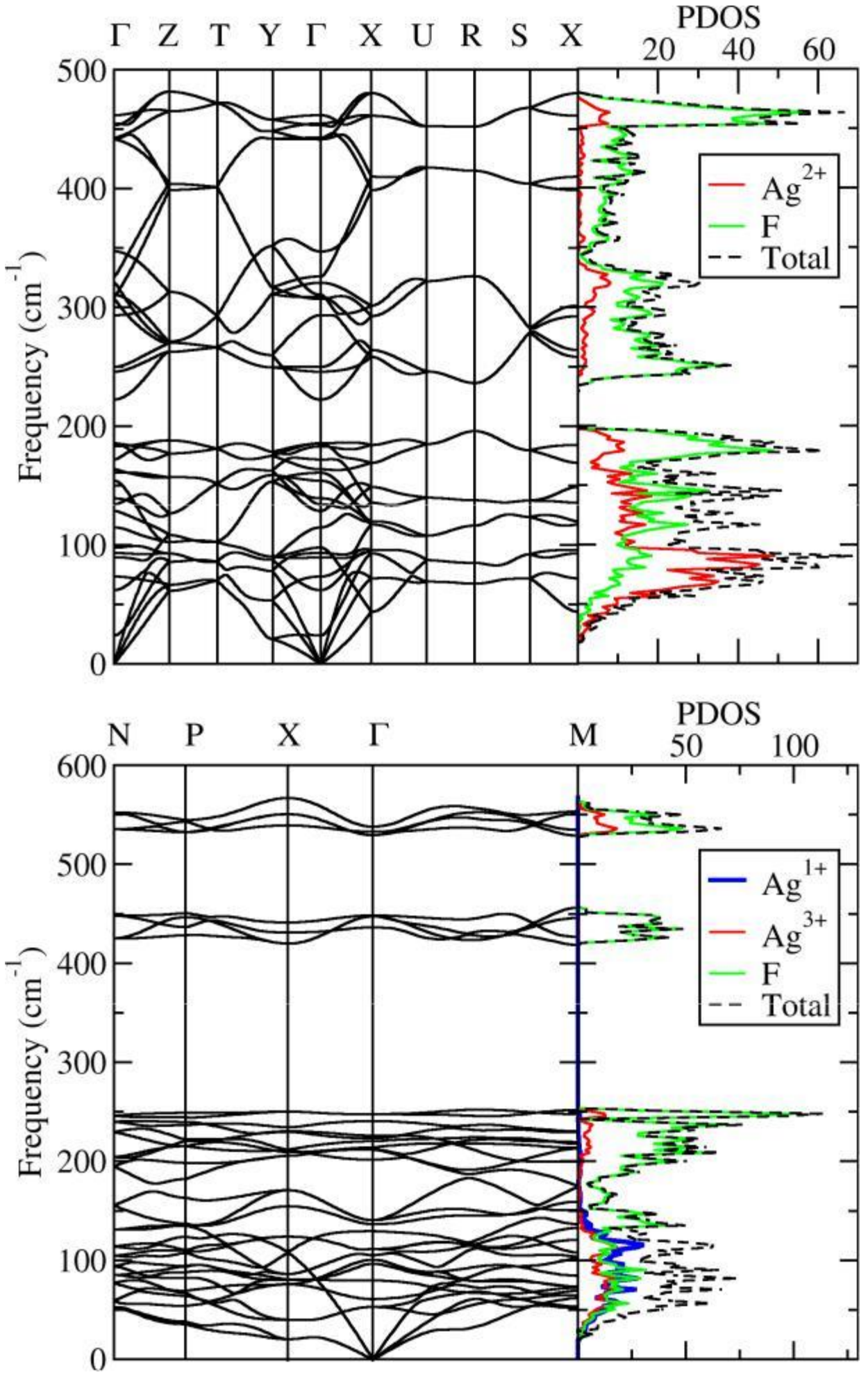

Fig. 5 Calculated phonon dispersion curves and phonon density of states of α- (top) and β -AgF2 (bottom). DFT+U results.

their strong coupling and consequently in appearance of one con-tinuous energy region between 225 and 500 cm $^{1}$. In β , on the other hand, the high-energy stretching modes are localized into two very narrow energy regions less than 50 cm $^{1}$ wide. They are centred at 440 and 540 cm $^{1}$, and belong to symmetric and asymmetric bond stretching vibrations within [$AgF_4$] units, re-spectively. They are completely separated from each other by *cca.* 80 cm $^{1}$ wide phonon bandgap as well as from the lower-energy region by 180 cm $^{1}$ wide bandgap. The lower-energy re-gion spreads over the energy range 0–250 cm $^{1}$. The PDOS reveals that the phonon spectrum of β is dominated by the contri-butions from fluorine atoms similarly as in α. On the other hand, noticeable differences can be observed in redistribution of the silver phonon states that is additionally related to presence of two distinct silver atoms. The square planar $Ag^{3+}$ silver contributes to all but the bands centered around 430 cm $^{1}$, while the square antiprismatic $Ag^{1+}$ silver contributes only to the lowest energy bands below 150 cm $^{1}$. The large separation of the symmetric and asymmetric [$Ag^{3+}F_4$] stretching vibrations (D = 80 cm $^{1}$) contrasts with the situation in the α phase, where the respective [$Ag^{2+}F_4$] vibrations are only *cca.* 10 cm $^{1}$ apart and centered around 450 cm $^{1}$. The Γ-point centered asymmetric [$Ag^{3+}F_4$] stretching vibrations appear at much higher energies (530–540 cm $^{1}$) in respect to the asymmetric [$Ag^{2+}F_4$] ones in a (450–460 cm $^{1}$). On the other hand, the symmetric [$Ag^{3+}F_4$] stretching vibrations appear at only slightly higher energies (447–450 cm $^{1}$) in respect to the symmetric [$Ag^{2+}F_4$] ones (440–445cm $^{1}$). The overall higher $Ag^{3+}$–F stretching frequencies are in line with formation of stronger covalent bonds in the β phase. Also, contribution of the square antiprismatic silver only to the lowest energy bands is in line with formation of closed shell $Ag^{1+}$ cation. The above described features of lattice dynamics of α and β phase manifest presence of chemically and electronically distinct lattices. Highly localized dispersionless bands in β manifest the presence of the electronically isolated [$Ag^{3+}F_4$] plaquettes and thus the lack of extended crystal network of covalent bonds. On the other hand, the highly dispersed and strongly coupled phonon bands in α manifest presence of the extended network of covalent $Ag^{2+}$–F bonds. One can observe that the largest dispersion of the bands in a takes place mostly along the Γ-X, Γ-Y , Γ-Z and

T-Y directions in the 1$^{st}$ Brillouin zone.

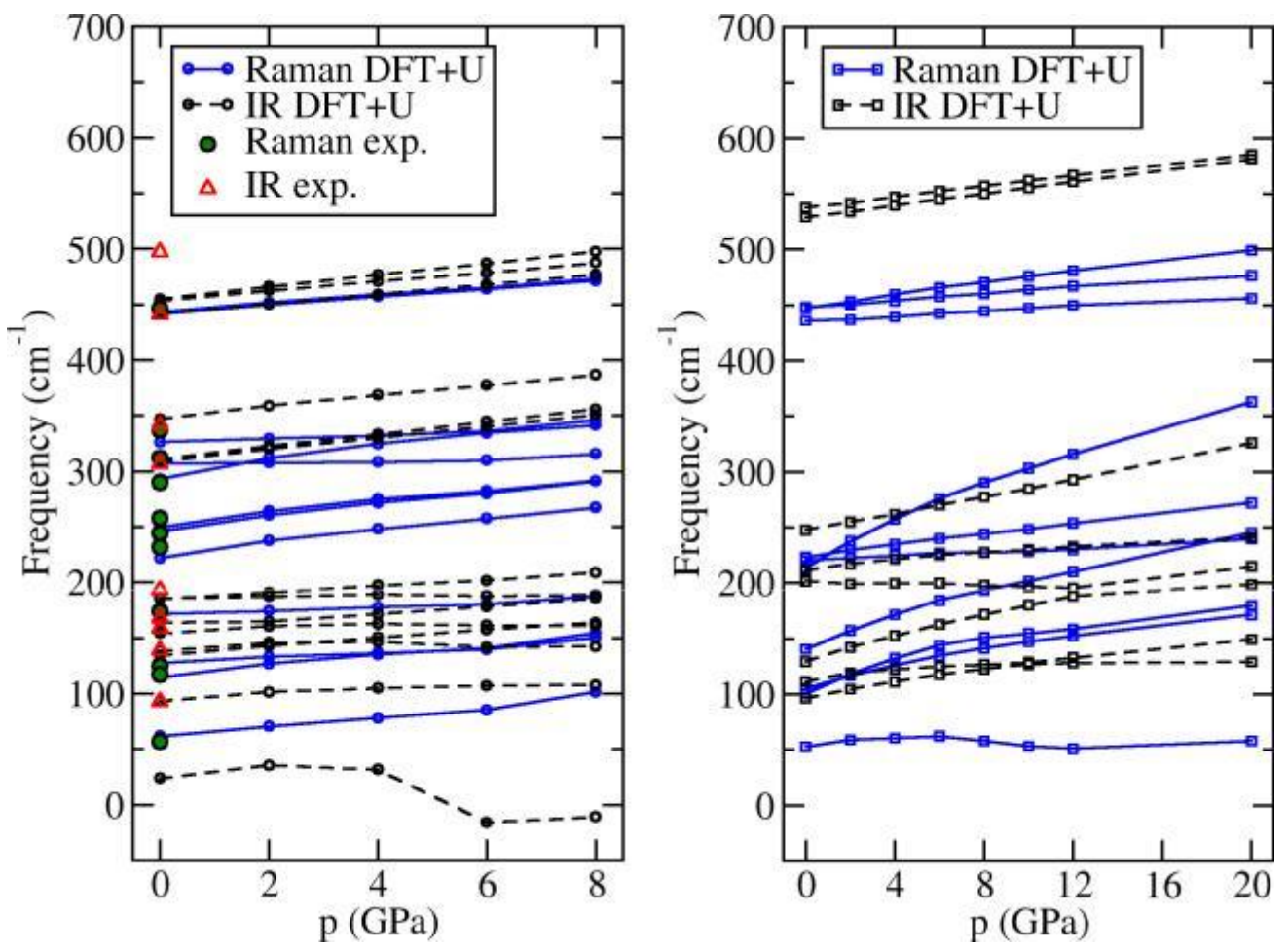

Fig. 6 Pressure-dependence of Raman and IR active frequencies calculated for α (left) and β (right) $AgF_2$ phase. DFT+U results.

Complete list of Γ-point frequencies of the α and the β phase at zero pressure is provided in Table 3 together with their symmetry and optical activity. In case of the α phase, the frequencies are compared with experimental IR and Raman data from literature and overall good agreement is reached in line with our previous findings. [1] In both $AgF_2$ phases, the Raman and IR frequencies increase monotonically and quite linearly with pressure in the entire pressure ranges considered within this study as shown in Figure 6. The only two exceptions are the lowest energy IR-active $B_{2u}$ mode in a and Raman-active $E_g$ mode in the β phase. The latter slightly softens above 6 GPa, but the structure remains dynamically stable even at 20 GPa. This contrasts with the situation in the α phase, where the lowest-energy $B_{2u}$ mode becomes dynamically unstable at 6 GPa (witnessed by negative energy values in Figure 6). Softening of this mode was used to explain the phase transition from the orthorhombic Pbca to the high-pressure $Pca2_1$ structure in the a phase. [8]

### 3.4 Phase diagram and thermodynamic stability

The comparison of the static energy dependence on volume and correction to vibrational zero-point energies (ZPE) is shown for both $AgF_2$ phases in left panel of Figure 2. The static energy minima are well distinguished and they have similar energy dependence on volume, while β has higher ground state energy than α at all calculated pressures. At zero pressure the difference amounts to 0.146 eV/$AgF_2$, which is close to the previously calculated DFT+U value of 0.17 eV/$AgF_2$. [6] Calculated ZPEs per one $AgF_2$ unit are also strikingly similar within the pressure range studied; they amount to 0.125 eV and 0.122 eV per one $AgF_2$ unit for α and β , respectively. Thus, inclusion of the ZPE to the internal energy does not affect the relative stability of both forms. The calculated ground state thermodynamic data are summarized in Table 4.

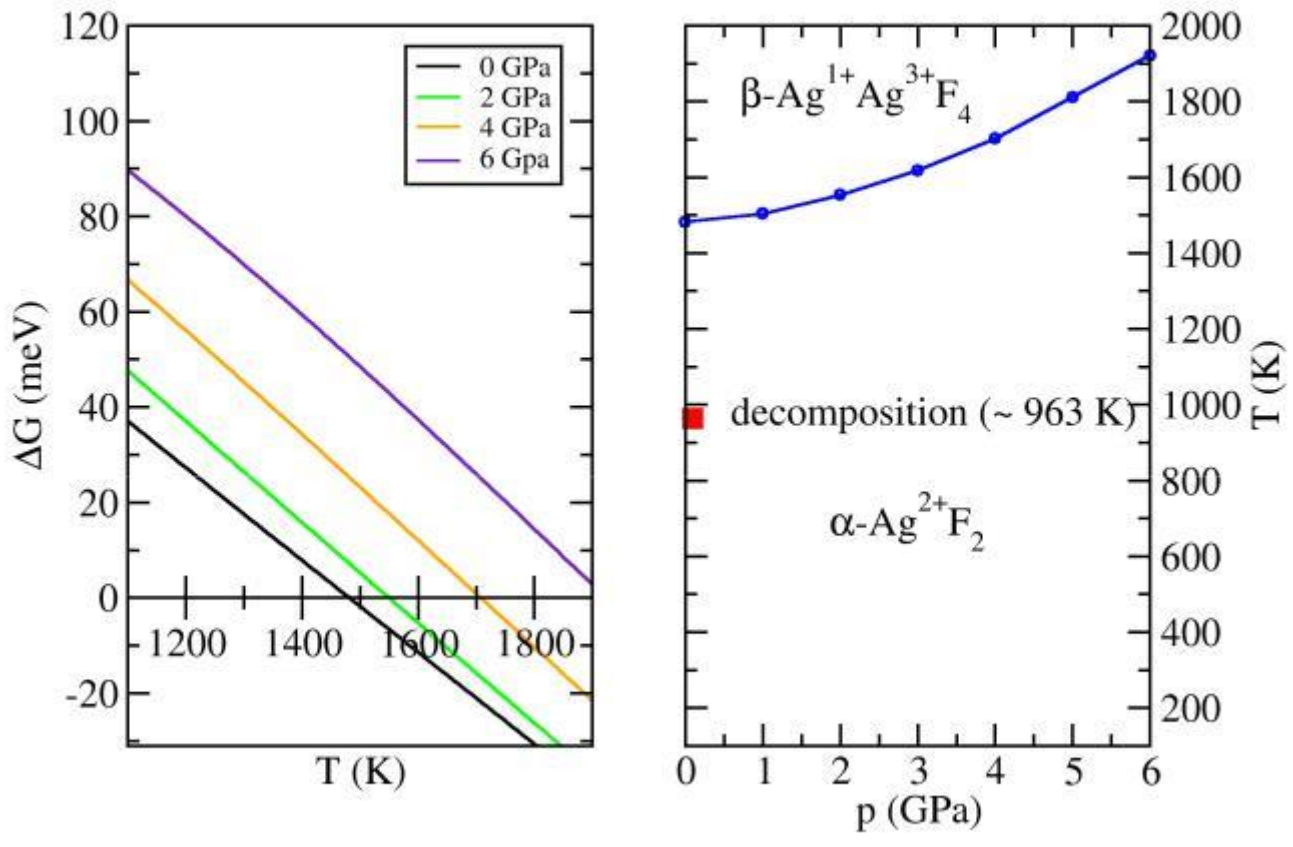

Fig. 7 The Gibbs energy isobars of thermodynamic equilibrium between the α and the β $AgF_2$ phase in the temperature range 1500—1900 K (left) and α–β phase diagram (right).

Impact of temperature on relative thermodynamic stability of the α and β phase is illustrated in Figure 7. The left panel shows relative Gibbs energy differences between α and β in function of temperature at constant pressure. The Gibbs energy isobars were evaluated at five pressure points within the region 0–8 GPa. They represent thermodynamic equilibria between α and β at constant pressure. The phase boundary between α and β was determined from sections of the isobar with the x-axis of the dia-gram. The obtained α– β phase diagram is visualized for the pressure range 0–6 GPa and temperature range 1500–1900 K in left panel of Figure 7. These calculations indicate that α is the thermodynamically stable phase in a broad field of temperature and pressure. β becomes thermodynamically preferred only at very high temperature, which additionally increases with pressure. At zero pressure, stability field of β over α extends above 1500 K, which, however, substantially exceeds the experimentally determined value of thermal decomposition temperature of the latter (963 K). [41] The fact that the β phase was observed in experiments at ambient conditions suggests that it is metastable with respect to a form at lower temperatures. This is in accordance with its dynamic (phonon) stability as computed here (Figure 5, bottom). Moreover, the experimentally observed mild exoergonic transformation to the α phase [10] agrees with the relative energies of both phases derived here from the first principles (0.14 eV corresponds to 13.5 kJ/mol).

Table 3 Calculated G-point frequencies of the α- and β -$AgF_2$. In case of a, available experimental data are included. Symmetry and optical activity (in brackets) of the normal modes is shown: IR – infrared active, R – Raman active, w – weak, sh – shoulder, m – medium intensity peak.

| a | | | | | | | | b | | | | | |
|---|---|---|---|---|---|---|---|---|---|---|---|---|---|
| # | Irr | f/cm$^{-1}$ Exp | f/cm$^{-1}$ DFT+U | # | Irr | f/cm$^{-1}$ Exp | f/cm$^{-1}$ DFT+U | # | Irr | f/cm$^{-1}$ DFT+U | # | Irr | f/cm$^{-1}$ DFT+U |
| 1 | B2u (IR) | | 24 | 18 | Ag (R) | 232-233 | 222 | 1 | Eg (R) | 53 | 18 | Eu (IR) | 248 |
| 2 | Ag (R) | 57 | 62 | 19 | B3g (R) | 244 | 246 | 2 | B1u | 61 | 19 | A1g (R) | 436 |
| 3 | Au | | 73 | 20 | B2g (R) | 258 | 250 | 3 | A2g | 80 | 20 | Eg (R) | 447 |
| 4 | B1u (IR) | 93w | 90 | 21 | B1g (R) | 290-291 | 293 | 4 | Eu (IR) | 96 | 21 | B2g (R) | 448 |
| 5 | B3u (IR) | 93w | 93 | 22 | Ag (R) | 311-312 | 307 | 5 | B1g (R) | 101 | 22 | Eu (IR) | 529 |
| 6 | Au | | 98 | 23 | B1u (IR) | 307m | 309 | 6 | Eg (R) | 105 | 23 | B1u | 533 |
| 7 | B1g (R) | 102-117 | 115 | 24 | B2u (IR) | 307m | 310 | 7 | Eu (IR) | 111 | 24 | A2u (IR) | 538 |
| 8 | B3g (R) | 125 | 127 | 25 | Au | | 321 | 8 | A2u(IR) | 129 | | | |
| 9 | B2u (IR) | 139sh | 134 | 26 | B1g (R) | 337 | 326 | 9 | B2u | 136 | | | |
| 10 | B1u (IR) | | 138 | 27 | B3u (IR) | 341sh | 347 | 10 | Eg (R) | 140 | | | |
| 11 | B3u (IR) | 159w | 154 | 28 | B3u (IR) | 441vs | 441 | 11 | A2u (IR) | 201 | | | |
| 12 | Au | | 161 | 29 | B3g (R) | | 442 | 12 | Eu (IR) | 212 | | | |
| 13 | B3u | 168m | 164 | 30 | B2g (R) | 446 | 443 | 13 | A2g | 214 | | | |
| 14 | B2g (R) | 162-174 | 172 | 31 | B1u (IR) | | 453 | 14 | A1g (R) | 220 | | | |
| 15 | Au | | 183 | 32 | B2u (IR) | | 455 | 15 | B2g (R) | 223 | | | |
| 16 | B1u (IR) | 193m | 185 | 33 | Au | | 462 | 16 | A1u | 226 | | | |
| 17 | B2u (IR) | 193m | 186 | | | | | 17 | B1u | 241 | | | |

Table 4 Ground-state energies ($E_{GS}$), zero-point energies (ZPE), relative ground state and free energies ($\Delta E_{GS}$, $\Delta E_{Free}$), the volumes per one $AgF_2$ unit (V), bulk moduli ($B_0$) and their pressure derivatives ($B'_0$) calculated for α- and β -$AgF_2$ at p,T = 0. $B_0$ and $B'_0$ were fitted to Birch-Murnaghan equation of state. [40] The energies are calculated per one $AgF_2$ unit.

| | $\Delta E_{GS}$ (eV) | $E_{GS}$ (eV) | ZPE (eV) | $\Delta E_{Free}$ (eV) | V (Å$^3$) | $B_0$ (GPa) | $B'_0$ (GPa) |
|---|---|---|---|---|---|---|---|
| a | 0 | -9.392 | 0.125 | 0 | 40.46 | 50 | 8.38 |
| b | 0.146 | -9.248 | 0.122 | 0.141 | 40.83 | 54 | 5.84 |

### 3.5 Comparison of β -$AgF_2$ with $NaCuO_2$ and AgO

One can compare β -$AgF_2$ and oxocuprate $NaCuO_2$, which has been studied as reference compound for the existence of $Cu^{3+}$ species in high-T oxocuprate superconductors. [14–16] Both compounds contain formally $d^8$ cation ($Ag^{3+}$ or $Cu^{3+}$) in square planar coordination with the ligands (respective building units are [$Ag^{3+}F_4$] or [$Cu^{3+}O_4$] plaquettes) and are diamagnetic insulators with the metal $Cu^{3+}$/$Ag^{3+}$ states positioned at lower energies relative to the p states of ligands. The insulating band gap of $NaCuO_2$ was measured by photoemission spectroscopy to be 1 – 2 eV, [15] which is comparable to bandgap of β from hybrid DFT (1.2 eV). The two compounds however differ in crystal structure and type of the counter cation that greatly define the details of electronic structure. The isolated [$Ag^{3+}F_4$] plaquettes in β - $AgF_2$ are contrasted with infinite [$Cu^{3+}O_2$] chains of edge-sharing [$Cu^{3+}O_4$] plaquettes. [42] Additionally, the valence and conduction bands in $NaCuO_2$ are constructed only from the $Cu^{3+}$ d and O p states, [16] while in case of β -$AgF_2$, the picture is additionally complicated by presence of the $Ag^{1+}$ d states just below the Fermi level (Figure 3, bottom). These additional states push the $Ag^{3+}$ d and F p states further below the Fermi level in respect to the $Cu^{3+}$ d and O p states in $NaCuO_2$. [16] These differences account for distinct character of the band gap in both systems.

It is also instructive to compare β -$AgF_2$ with its oxygen counterpart $Ag^{1+}Ag^{3+}O_2$. In this case the two compounds are truly isoelectronic with comparable hybrid DFT values of insulating bandgap ( 1 eV), [22,23,43] which in the case of AgO agree very well with the experimental value of 1.0–1.1 eV. [44,45] In the monoclinic AgO, the $Ag^{3+}$ cations form two-dimensional infinite puckered layers of corner-sharing [$Ag^{3+}O_4$] plaquettes. [17] This contrasts the isolated [$Ag^{3+}F_4$] plaquettes in β -$AgF_2$ but is reminiscent of the puckered layers in a-$AgF_2$. Concerning the electronic DOS, the highly localized character of the $Ag^{1+}$ and $Ag^{3+}$ states in β - $AgF_2$ (Figure 3, bottom and S5 in ESI) is quite distinct also from the highly dispersed $Ag^{1+}$ and $Ag^{3+}$ d states in AgO where they additionally strongly overlap with each other and O p states. [22] Thus, charge disproportionation in β -$AgF_2$ is more pronounced relative to AgO.

## 4 Conclusions

Crystal and electronic structure, lattice dynamics and thermodynamic stability of little known mixed-valent diamagnetic $Ag^{1+}Ag^{3+}F_4$ β form of $AgF_2$ was thoroughly examined for the first time and compared with well known antiferromagnetic $Ag^{2+}F$ α form within the framework of Density Functional Theory based methods, phonon direct method and quasiharmonic approximation. Results of a detailed analysis of lattice dynamics and impact of temperature on stability of the a phase were also for the first time presented. The modelled $KBrF_4$ type crystal structure of β is comparable to $NaAgF_4$ in line with well known similarities of $Na^+$ and $Ag^+$ cations, and its tetragonal c/a ratio fits ideally within the observed trend among the ternary metal fluorides $M'MF_4$ with coinage (M) and alkali metals (M'). Crystal structure of both, α and β , are structurally closely related and can be derived from the common fluorite parent prototype via distinct order parameter X5+ and L3 + /L3 , respectively. Their volumes are also very close, the volume of β is only sightly larger (1%), and both exhibit comparable compressibility ($B^{\alpha}_0$ = 50 GPa and $B\beta_0$ = 54 GPa), which is simultaneously considerably higher then for their lighter copper counterpart $CuF_2$. Electronic struc-ture calculations revealed highly localized $Ag^{1+}$ and $Ag^{3+}$ states in β relative to highly dispersed $Ag^{2+}$ ones in α. Both phases are insulators, the bandgap of β being by factor of 2 lower at all considered pressures, as revealed by DFT+U results (at zero pressure $E_{gap}{}^{\alpha}$ = 1.167eV and $E_{gap}{}^{\beta}$ = 0.61 eV). The hybrid DFT bandgaps were computed to be by factor of 2 wider. Phonon densities of states reveal pronounced differences related to pres-ence of distinct Ag species in each case. In α, the stretching and bending modes are highly coupled and the overall phonon bands are greatly dispersed. On the other hand, phonon structure of β is characteristic of highly localized bands, the symmetric and asymmetric $[Ag^{3+}F_4]$ vibrations are completely decoupled from each other as well as from the lower-energy bending and lattice modes. The considerably increased frequencies of the asymmetric $[Ag^{3+}F_4]$ modes relative to those of $[Ag^{2+}F_4]$, confirm presence of increased covalency of chemical bonds in β relative to α. In both phases, the IR and Raman active modes increase monotonically with pressure. The only exception is the lowest-energy IR active $B_{2u}$ mode in α and the lowest-energy Raman $E_g$ mode in β. It was show that the former mediates the phase transition to ferroelectric $Pca2_1$ structure in α. [8] On the other hand, the latter softens only slightly and β remains dynamically stable at all calcu-lated pressures up to 20 GPa. The α phase is thermodynamically preferred over β also at high pressures and temperatures. β phase cannot be reached experimentally in close-to-equilibrium condi-tions with α: its calculated temperature stability region lies far above the thermal decomposition temperature of alpha phase in the p–T diagram. Computed results are in line with experimental observation: β phase may be obtained exclusively as a metastable species when starting from $Ag^{1+}$ and $Ag^{3+}F_4$ precursors. [41] Fi-nally, the insulating hybrid DFT bandgap of β is comparable with bandgaps of related mixed-valent $Na^{1+}Ag^{3+}O_2$ and $Ag^{1+}Ag^{3+}O_2$, and electronic disproportionation is more pronounced in β -$AgF_2$ relative to AgO.

## Acknowledgements

K.T. and M.D. acknowledge the European Regional Development Fund, Research and Innovation Operational Program (project No. ITMS2014+: 313011W085), the Slovak Research and Develop-ment Agency (grant No. APVV-18-0168) and Scientific Grant Agency of the Slovak Republic (grant No. VG 1/0223/19). WG acknowledges the Polish National Science Center (NCN) for the Beethoven project (2016/23/G/ST5/04320). The research was carried out using machines of the Interdisciplinary Centre for Mathematical and Computational Modelling (ICM), University of Warsaw under grant ADVANCE++ (no. GA76-19) and Aurel supercomputing infrastructure in CC of Slovak Academy of Sci-ences acquired in projects ITMS 26230120002 and 26210120002 funded by ERDF.

## Notes and references

# Supporting Material (ESI) to

## Comparative computational study of antiferromagnetic and mixed-valent diamagnetic phase of $AgF_2$: crystal, electronic and phonon structure and p-T phase diagram

Kamil Tokár,[a;b] Mariana Derzsi,[a;c] and Wojciech Grochala[c]

[a]*Advanced Technologies Research Institute, Faculty of Materials Science and Technologyin Trnava, Slovak University of Technology in Bratislava, J. Bottu 25, 917 24 Trnava, Slovakia. E-mail: mariana.derzsi@stuba.sk*

[b]*Institute of Physics, Slovak Academy of Sciences, Dúbravská cesta 9, 845 11Bratislava, Slovakia.*

[c]*Center of New Technologies, University of Warsaw, Żwirki i Wigury 93, 02089 Warsaw, Poland.*

**Table S1.** Comparison of α-$AgF_2$ and β-$AgF_2$ lattice parameters with function of pressure. **DFT+U** results.

| | **α-$AgF_2$** | | | **β-$AgF_2$** | |
|---|---|---|---|---|---|
| **P (GPa)** | **a(Å)** | **b (Å)** | **c(Å)** | **a (Å)** | **c (Å )** |
| 0 | 5.498 | 5.824 | 5.055 | 5.524 | 10.704 |
| 2 | 5.466 | 5.730 | 5.005 | 5.449 | 10.653 |
| 4 | 5.445 | 5.650 | 4.965 | 5.387 | 10.610 |
| 6 | 5.4285 | 5.610 | 4.900 | 5.332 | 10.575 |
| 8 | 5.420 | 5.570 | 4.844 | 5.286 | 10.543 |
| 10 | NA | NA | NA | 5.244 | 10.517 |
| 20 | NA | NA | NA | 5.077 | 10.423 |

**Fig S1**. Electronic density of states for α-$AgF_2$ (top) and β-$AgF_2$ phase (bottom) at 0 GPa. The Fermi level is set to 0 eV. **Hybrid DFT** results.

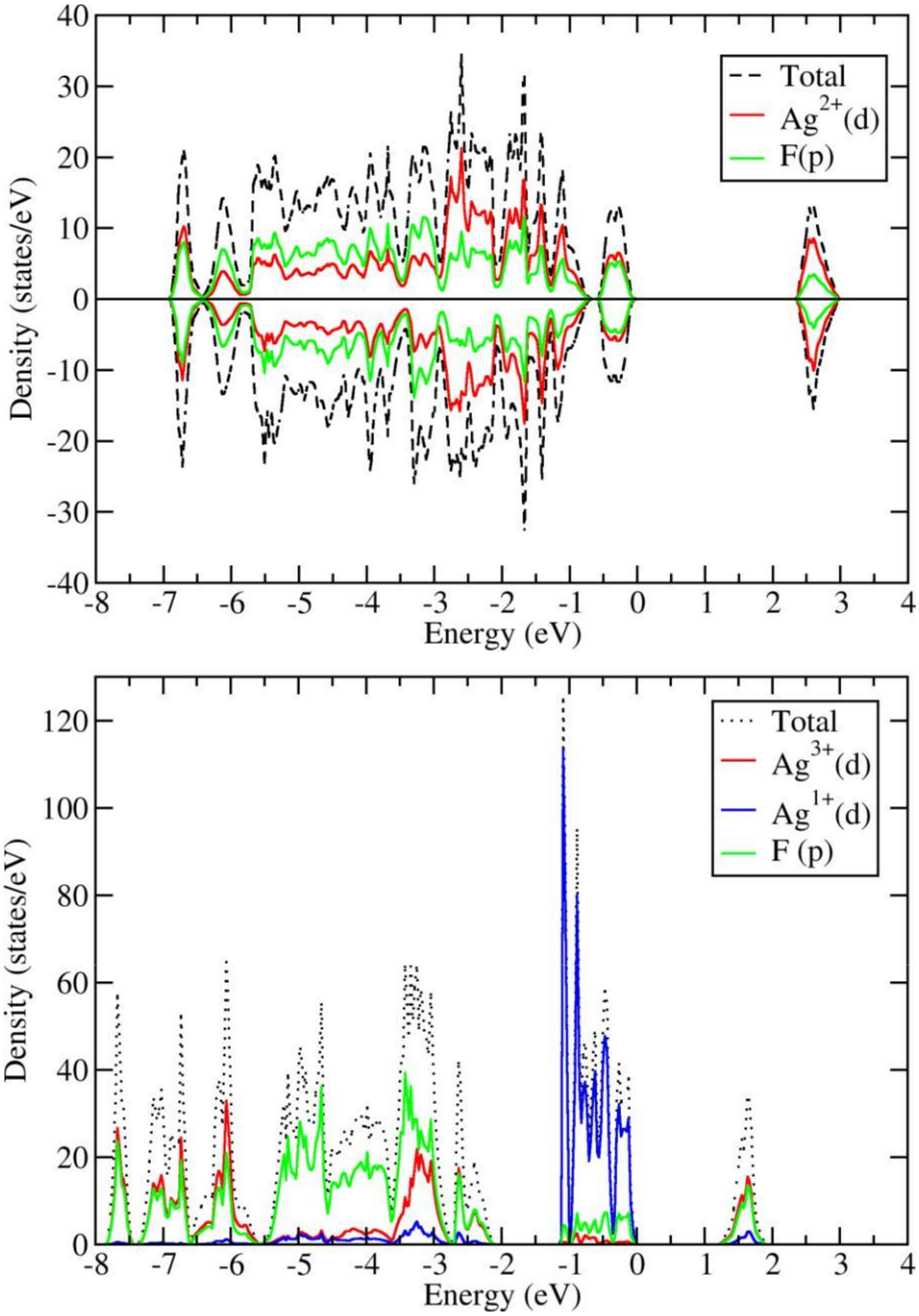

**Fig S2.** Electronic density of states for **α**-$AgF_2$ phase in pressure range 0 - 6 Gpa. The Fermi level is set to 0 eV. **DFT+U** results.

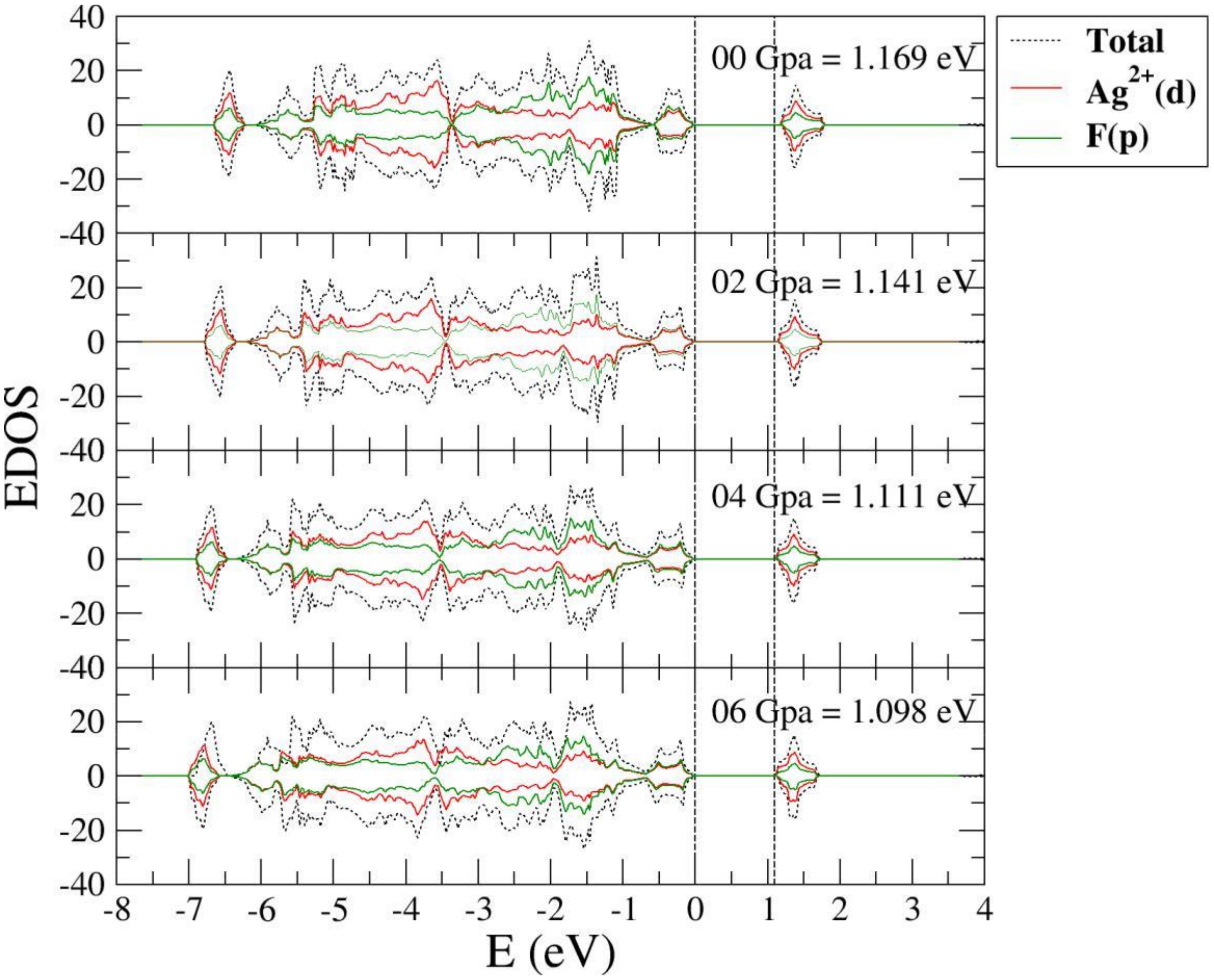

**Fig S3.** Electronic density of states for **β**-$AgF_2$ phase in pressure range 0 - 6 Gpa. The Fermi level is set to 0 eV. **DFT +U** results.

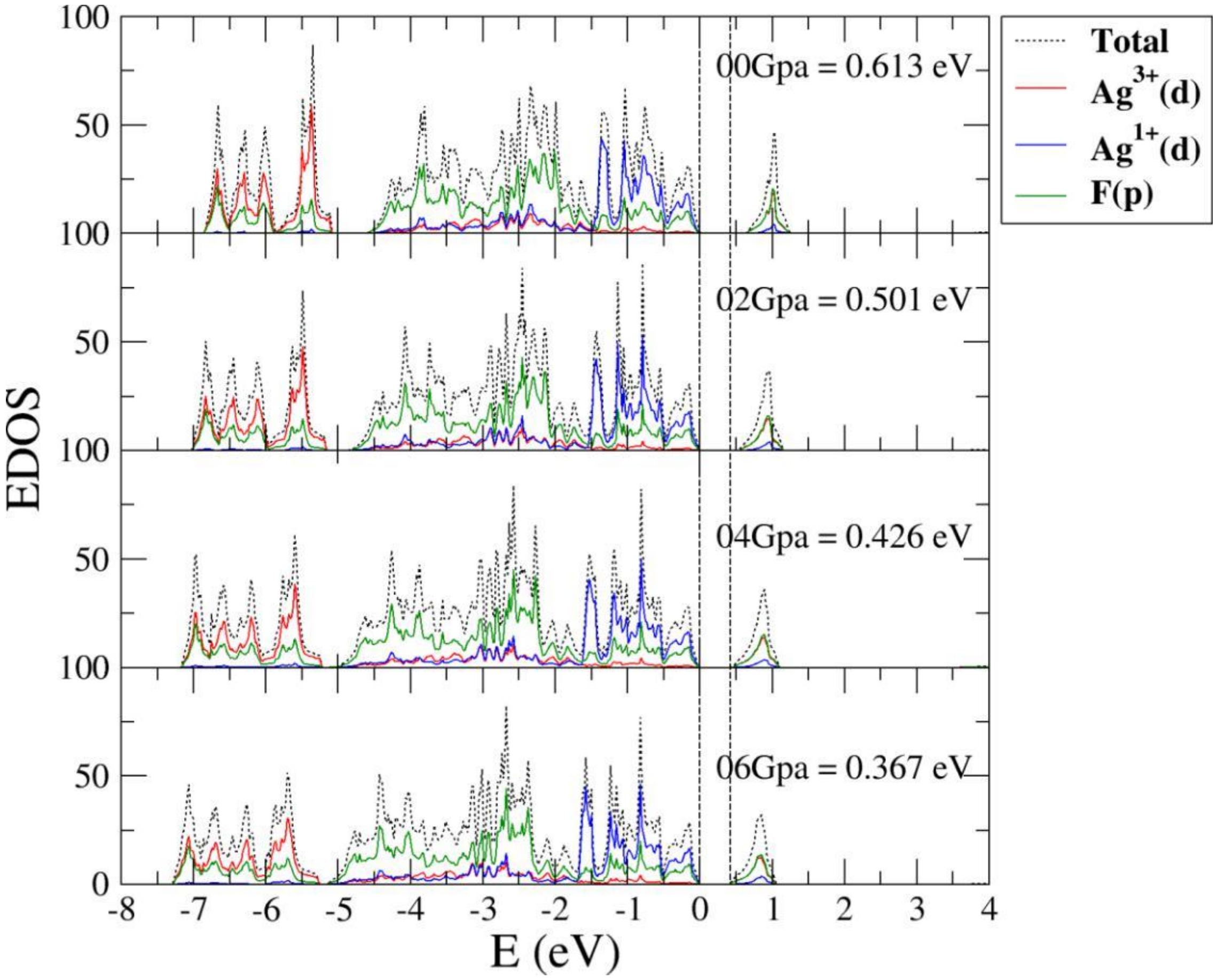

**Fig S4.** Electronic density of states for **α**-$AgF_2$ phase in pressure range 0 - 6 Gpa. The Fermi level is set to 0 eV. **Hybrid DFT** results.

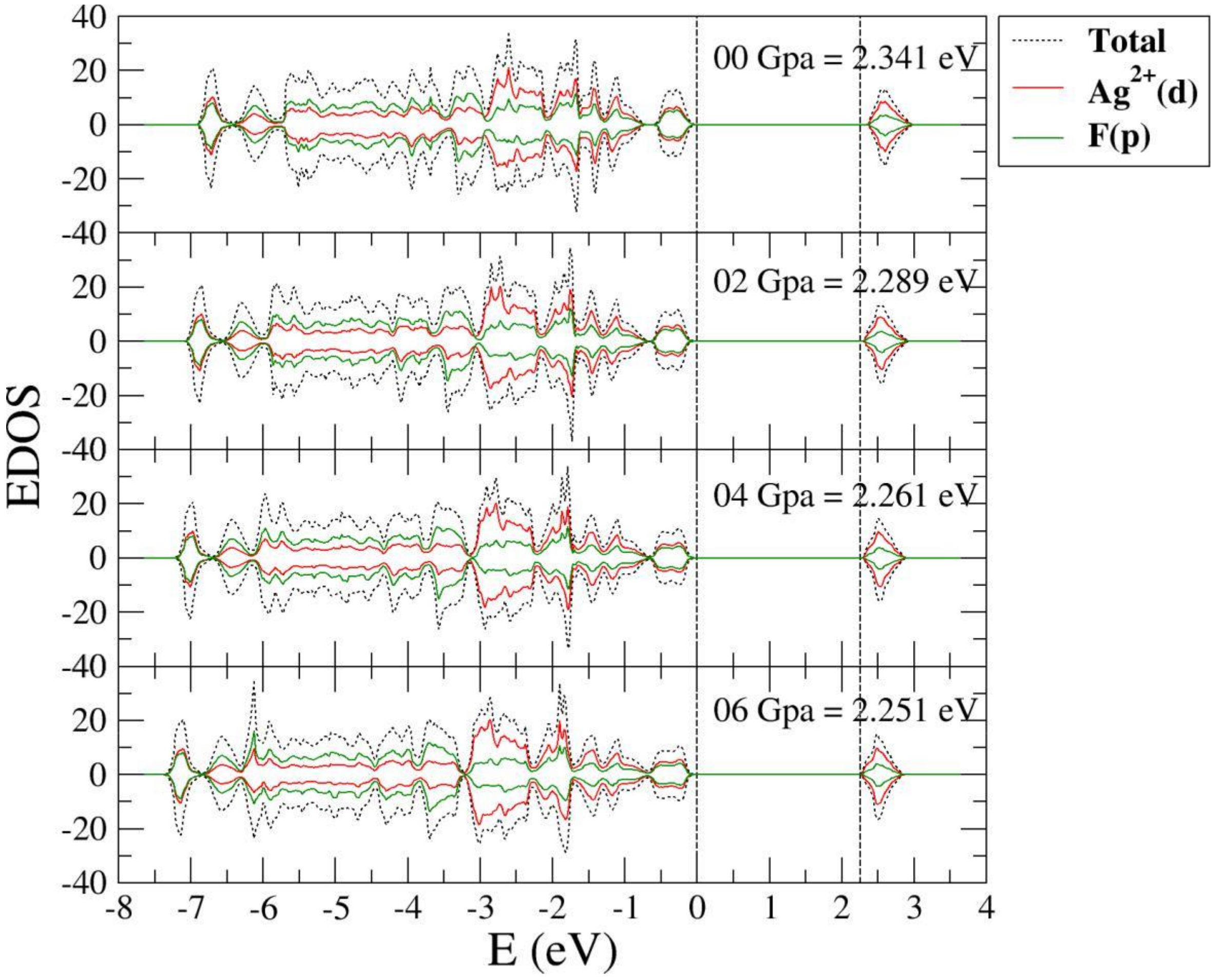

**Fig S5.** Electronic density of states for **β**-$AgF_2$ phase in pressure range 0 - 6 Gpa. The Fermi level is set to 0 eV. **Hybrid DFT** results.

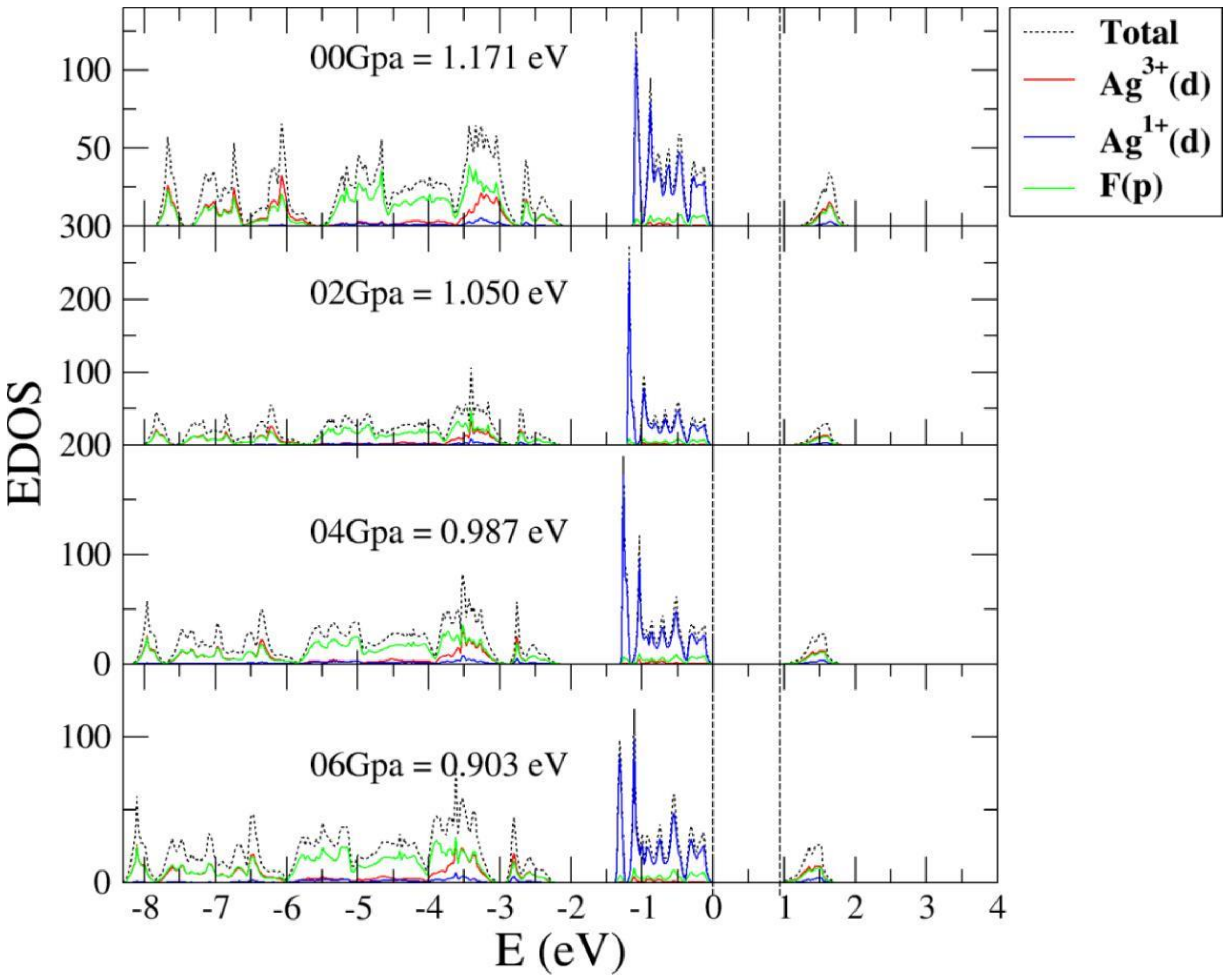